# Do Elementary Particles Have an Objective Existence?

Bilha Nissenson

**22 Nov 07**

The formulation of quantum theory does not comply with the notion of objective existence of elementary particles.  Objective existence independent of observation implies the distinguishability of elementary particles. In other words: If elementary particles have an objective existence independent of observations, then they are distinguishable. Or if elementary particles are indistinguishable then matter cannot have existence independent of our observation.
This paper presents a simple deduction of the above statements, their compatibility with quantum theory, an example of quantum uniqueness situation and a suggested experiment. The conclusion is a short discussion about the redundancy of such phenomena.

## Introduction

*"…if we consider any concept (e.g. apples), then this concept contains nothing in it that would permit us to distinguish one apple from another."*
David Bohm, "Causality & Chance in Modern Physics" published in 1957 [1]. Bohm continues his thought; *"For even the modes of being of the individual atoms, electrons, protons, etc., inside the apple are in turn determined by infinity of complex processes in their substructures and background".*   And he concludes that …*"such a thing can and indeed must be unique: i.e. not completely identical with any other thing in the universe, however similar the two things may be."*
In 1964 John Bell [2] drew the attention of physicists to the extra ordinary features of entanglements; quantum theory describes a pair of entangled objects as a single global quantum system. He demonstrated that there is no way to understand entanglements in the frame of the usual ideas of a physical reality localized is space time and obeying causality. This is in contrast to Einstein Podolsky and Rosen, (EPR) expectations as formulated in their paper [3]. Bell's work offered the possibility to determine experimentally whether or not Einstein's ideas are kept.
In 1984 Alan Aspect et Al.  performed an experiment, based on Bell's theorem, and revealed that;
1. Measuring one of the entangled particles determines with certainty the state of the other (like EPR predicts).
2. The entanglements cannot be understood as usual correlations, whose interpretation relies on the existence of common properties, originating in a common preparation, and remaining attached to each individual object after separation, as components of their physical reality. [4]
The result of this experiment could be explained by assuming that each of the photons coming out of the entangled pair situation had its polarization direction regardless of observations or measurements, as Einstein suggested in the EPR article [3].

## Simple deduction

We are used to phenomena around us being unique, we know that each snow flake has a unique shape, so does each grain of sand, and those are just inanimate phenomena. The uniqueness is formed as a result of the unique conditions at the time of the flake (grain, etc) being formed, i.e. the initial conditions. For a snow flake we define t=0 as the time when 2 molecules of water formed the first snow bond between them, t=Δt is when a third molecule joined in, and so on. So the shape of a snow flake depends on the exact situation of each of the water molecules joining together to form it. The outcome of the assembly of molecules and the unique set of conditions that is present at the time of forming each bond result in the uniqueness of each snow flake. [5]

The forming of a snow flake is a classical phenomenon, elementary particles are quantum phenomena, Let us consider the assumption that quantum phenomena such as electrons, protons, or photons, have an objective existence and evaluate the consequences.

Time t=0 for protons or electrons happened very long ago, as long as our universe exists, we will mark t= Δt as first event; an interaction [6] with another particle, time t=2 Δt is the next event and so on. This way we can define for each objectively existing entity a distinguishable path in space time, marked by a unique sequence of events [7].

Exchanging 2 such electrons means exchanging the initial conditions, this for objective existing entities, will result in chaotic diversification in the same described in Lorenz's "simulated weather" [8].

Pauli principle is not contradicted here;

Putting it in Bohm's words; [1],

*"According to Pauli Exclusion Principle, any two electrons are said to be "identical". This conclusion follows from the fact that within the framework of the current quantum theory there can be no property by which they could be distinguished. On the other hand, the conclusion that they are completely identical in all respects follows only if we accept the assumption of the usual interpretation of quantum theory that the present general form of the theory will persist in every domain that will ever be investigated. If we do not make this assumption, then it is evidently always possible to suppose that distinction between electrons can arise at deeper levels."*

## Compatibility with quantum theory

A quantum state is a wave function, or an eigenvector that solves the Schrödinger equation. It describes mathematically all possible out comes of a quantum system when a measurement will occur. Each outcome is accompanied by the probability of that outcome to occur. We assume that the wave function contains all the accessible information relating to the system under description. This means that two systems with identical state can have different outcome upon measurement (Bohm, Schiff, and others [9], [10]) therefore the quantum state or wave function of a system is not in one to one correspondence with the actual behavior of the system. When two systems have the same wave function we can only say that they have the same range of potentialities within them, and the same probability for a certain outcome, but we cannot say if they will have the same outcome upon measurement [9].

We note here that Bohr suggestion rejected this interpretation which gives independent reality to whatever is supposed to be represented by the wave function. His interpretation is that the quantum state is a mathematical entity which gets meaning only after a measurement has been done. This totally rejects any objective property of elementary particles.

In their experiments to test Bell's theorem [2], A. Aspect et Al [3] demonstrated EPR-type correlations between entangled pairs of photons, while in total retaining quantum predictions. These experiments, demonstrate that particles, in that case photons, had with certainty some property (direction of polarization), before a measurement was done. On the other hand quantum statistical predictions were confirmed as well, rejecting hidden variables explanation for the phenomena.

It seems that the wave function solving the Schrödinger equation is not be a complete description of reality; for example, the coordinate parameter, as discussed above, does not enter the wave function (initial conditions). In the above textbooks [9] [10], the most complex solution for the Schrödinger, is demonstrated for an Hydrogen Atom. This differential equation can be solved analytically for the case of a 2 body problem only. It is not mentioned that the solution is given for an isolated Hydrogen Atom existing without history and background. Constructing a truly isolated system is impossible; we construct a completely isolated system only mentally, in thought experiments. We can approach such a system with very hard work, for a very short time, but these systems do not occur naturally. If there was truly isolated matter in our universe we would not be able to notice it, it would be transparent in our world. For all practical purposes, solving as if the system is isolated is good enough, but when we try to probe further into understanding; "What are the elementary particles?" and the subtle influences of; locality, gravity, relativity, etc. we should consider a reevaluation of our approximations. Indistinguishability is the core of quantum physics, but if particles have properties regardless of observations then they will have initial conditions that we must consider. If we disregard them, quantum mechanics might be reduced to a statistical mathematical tool that predicts experiments outcomes.

**A quantum example for uniqueness of position**

In Quasicrystals atoms are joined together in a long range order. One can visualize this order as Penrose tiling, where a shifted copy will never match exactly with its original. In such an order each intersection, which in quasicrystals marks the position of an atom, is a unique position upon an infinite grid, in that sense each atom have a unique position and is distinguishable, in that sense. Atoms that form quasicrystals, (molecules), are quite heavy, and include at least thousands of elementary particles.

In their experiment, M. Torres and Al [11] have found out that the same order can be demonstrated for the electrons:

 *"The spacing between the wave peaks (of the electron waves) was not constant, as in a periodic wave, but <u>varied quasiperiodically</u> between two values, which were related to the spacing in the pattern on the pan's bottom."*

Experiments revealing the exact circumstances for which a diffraction of quasicrystals disappear, by marking (with radioactive isotope etc.) very few atoms of the grid, might reveal information that will enable further understanding of the role of the unique position of each atom in the quasiperiodic grid. For observable results, doing the diffraction on the molecules as is done now would not be enough, what is needed is to get a refraction off the wave peaks as described in [11] but then how could we mark the electrons?

**Conclusion or getting something out of nothing**

It is plausible to dismiss or take for granted, the initial conditions of elementary particles that make them distinguishable. We pay no attention to the uniqueness of snow flakes and grains of sand. We consider all the diversification around us as part of an unavoidable noise that burdens our calculations.

We think that natural phenomena are the consequences of random incidents that accumulate in very long periods of time, by chance to create the world we see. But what is chance?

Henri Poincare in *Science & Method* writes (1907): [12]

"*…a very small cause which escapes our notice determines a considerable effect that we cannot fail to see, and then we say that the effect is chance …*" These words have been proved and demonstrated in a mathematical form by Lorenz and others seventy years later [8]. '*Small causes*' can be identified as the *events* described above (determining the uniqueness of an individual particle path). If particles have an objective existence, then these exact paths are the '*very small causes …that determines a considerable effect*', or if particles do not have an objective existence then we have to define again what we mean when we say 'chance'.

When a very large numbers of particles interact over a long period of time, humble differences add up. For example, there may have been a time before atoms, and molecules existed, it is conceivable these entities came about with the evolution of the universe over very long periods of time; we should consider that this slow evolution did not stop.

I will conclude with Bohm's [1] words:

"*The uniqueness of each thing at each instant of time is reflected… by the inexhaustibility of qualities that are to be found in nature.*"

---